\documentclass[conference]{IEEEtran}
\usepackage[utf8]{inputenc}
\usepackage[T1]{fontenc}
\usepackage{microtype}
\usepackage{graphicx}
\usepackage{dblfloatfix}
\usepackage{filecontents}
\usepackage[noadjust]{cite}
\usepackage[hidelinks, bookmarks=false, draft]{hyperref}
\usepackage[keeplastbox]{flushend}
\usepackage{amsmath}
\usepackage{url}
\usepackage{multirow}

\begin{document}

\title{Video as a By-Product of Digital Prototyping: \\Capturing the Dynamic 
Aspect of Interaction}

\author{
	\IEEEauthorblockN{
		Oliver Karras,
		Carolin Unger-Windeler, 
		Lennart Glauer, and
		Kurt Schneider
	}
	\IEEEauthorblockA{
		Software Engineering Group\\ 
		Leibniz Universität Hannover\\
		30167 Hannover, Germany\\ 
		Email: 
		\{oliver.karras, carolin.unger-windeler, 
		kurt.schneider\}@inf.uni-hannover.de,
		 lennart.glauer@stud.uni-hannover.de
	}
}
\maketitle

\begin{abstract}
Requirements engineering provides several practices to analyze how a user 
wants to interact with a future software. Mockups, prototypes, and scenarios 
are suitable to understand usability issues and user requirements early. 
Nevertheless, users are often dissatisfied with the usability of a resulting 
software. Apparently, previously explored information was lost or no 
longer accessible during the development phase.

Scenarios are one effective practice to describe behavior. However, they are 
commonly notated in natural language which is often improper to capture 
and communicate interaction knowledge comprehensible to developers and users. 
The dynamic aspect of interaction is lost if only static descriptions are used.

Digital prototyping enables the creation of interactive prototypes by adding 
responsive controls to hand- or digitally drawn mockups. We propose to capture 
the events of these controls to obtain a representation of the interaction. 
From this data, we generate videos, which demonstrate interaction 
sequences, as additional support for textual scenarios.

Variants of scenarios can be created by modifying the captured event 
sequences and mockups. Any change is unproblematic since videos only need to be 
regenerated. Thus, we achieve video as a by-product of digital prototyping. 
This reduces the effort compared to video recording such as screencasts. A 
first evaluation showed that such a generated video supports a faster 
understanding of a textual scenario compared to static mockups.

\end{abstract}


\IEEEpeerreviewmaketitle

\section{Introduction}
The most important requirements engineering goals are to create a 
\textit{shared understanding} between a project team and its stakeholders as 
well as \textit{specification quality} of requirements \cite{Fricker.2015c, 
Glinz.2015}. Requirements engineers need to achieve these objectives to bridge 
the communication gap between stakeholders and developers 
\cite{Gulliksen.2003}. This communication gap is critical since it may lead to 
unfulfilled customer expectations or communication of incorrect, unclear, 
ambiguous and non-verifiable requirements \cite{Bjarnason.2011}. This 
miscommunication threatens to achieve the software product goal 
\textit{usability} \cite{ISOIEC.1998} consisting of the sub-goals 
\textit{productivity} (in terms of efficiency), \textit{effectiveness}, and 
\textit{satisfaction} \cite{Fricker.2015c}. Thus, requirements engineering 
success is important to develop a software with satisfying usability.

Fricker et al. \cite{Fricker.2015c} identified \textit{scenarios}, which are 
exemplary sequences of system usage \cite{Alexander.2005}, as one 
of three practices that correlate with requirements engineering success. 
Scenarios support to achieve the most important requirements engineering goals 
and thus bridge the communication gap. They substantiate the functionality of a 
system and enable users to judge whether they presume to be able to use the 
system and whether they like it. Natural language is the most common notation 
to document scenarios \cite{Fricker.2015c, Alexander.2005}, for example in the 
format of the use case template \cite{Cockburn.2006}. According to Ambler 
\cite{Ambler.2002}, however, textual representations are the worst 
documentation option regarding communication. Smoots et al. \cite{Smoots.2016} 
support Ambler's perspective \cite{Ambler.2002} by emphasizing that exchanging 
requirements as textual descriptions can produce miscommunication. Fricker and 
Glinz \cite{Fricker.2010} evaluated the practice of handing-off written 
specifications and showed that its impact on the understanding of requirements 
is troublesome.

In contrast, videos are known to be the best documentation option for 
communication since several years \cite{Ambler.2002}. Karras et al. 
\cite{Karras.2016b} highlight the benefit of videos to capture verbal and 
non-verbal information comprehensively. Jirotka and Luff \cite{Jirotka.2006}, 
as well as Fricker et al. \cite{Fricker.2015}, emphasize the usefulness of 
videos as a stable reference for post-processing work. Developers can retrieve 
more details from a video than from any written documentation 
\cite{DeMarco.1990}. According to Brill et al. \cite{Brill.2010}, videos are 
appreciated for communication due to their richness and concreteness compared 
to text, which is perceived to be more precise but also more abstract.

Despite all these advantages, videos are not yet an established part of 
requirements engineering due to their high production effort 
\cite{Karras.2016}. Gulliksen and Lantz \cite{Gulliksen.2003} propose to use 
prototypes, mockups, and videos to support communication between a project team 
and its stakeholders. Prototyping is a highly valued technique to analyze 
scenarios about how users want to interact with a future software 
\cite{Rouibah.2009}. However, tools and methods are needed to integrate videos 
in existing activities \cite{Gulliksen.2003}. This need is supported by Carter 
and Karatsolis who suggested that ``research into a different set of tools 
aimed at capturing requirements and design activities, analyzing these records, 
and then producing effective clips might be a valuable investment'' \cite[p. 
4]{Carter.2009}.

According to this statement, we propose an approach of video as a 
by-product of digital prototyping to specify and document scenarios. We obtain 
an easy-to-modify and always repeatedly playable representation of interaction 
by capturing events of responsive controls of digital mockups. Thus, we can 
generate videos, demonstrating the dynamic aspect of interaction, as 
additional support for textual descriptions.

The contribution of this paper is an approach consisting of concepts to 
integrate video as a by-product of digital prototyping. We can reduce the 
effort of video production and support communication with a more suitable 
documentation option. Changes and variants of scenarios are no problems since 
videos can be easily regenerated from modified event sequences and mockups. We 
implemented the concepts of our approach in a prototypical software tool called 
\textit{Mockup Recorder}. In a first evaluation, we showed that such a 
generated video of our prototype supports a faster understanding of a textual 
scenario compared to static mockups.

The structure of the paper: Section \ref{ch:related_work} discusses 
related work. We describe our approach with its concepts in section 
\ref{ch:approach}. In section \ref{ch:evaluation}, we report our evaluation 
and its findings, which we discuss in section \ref{ch:discussion}. 
Section \ref{ch:conclusion} concludes the paper.

\section{Related Work}
\label{ch:related_work}
Several researchers already investigated the use of videos to enrich and 
document scenarios of how to use a software.

Mackay et al. \cite{Mackay.2000} used videos in their design process. In each 
phase, more detailed videos were created that demonstrate scenarios 
of using the system under development. The videos of a previous phase 
were input for the next one to bridge the gap between abstraction and detail. 
Zachos et al. \cite{Zachos.2005} developed the ART-SCENE tool to provide 
stakeholders more recognition cues for discovering requirements. They enhanced 
textual scenarios with rich media such as video and audio. Thus, they could 
describe the environment and other information of a system, which a textual 
description would have kept tacit.
The Software Cinema System of Creighton et al. \cite{Creighton.2006, 
Creighton.2006b} used videos which describe as-is and visionary scenarios of a 
system. The authors combined these videos with different types of artifacts as 
input to create hybrid videos consisting of Unified Modeling Language models 
and recordings from enacting a scenario as output. We follow their line of 
thought but emphasize and evaluate the use of video as a by-product of
digital prototyping to capture the dynamic aspect of interaction.
Broll et al. \cite{Broll.2007} reported on a methodological experiment that 
used videos to visualize concrete usage scenarios of a system under 
development. These videos were input for focus group discussions for 
requirements elicitation.
Maiden et al. \cite{Maiden.2007} investigated the effectiveness of different 
scenario forms and usages on requirements discovery. During workshops, 
stakeholders walked through scenarios which were presented in a textual and 
visual form. Their results reveal quantitative and qualitative differences in 
discovered requirements due to the presented scenario form.
Bruegge et al. \cite{Bruegge.2008} used video techniques to define 
requirements in a large-scale educational student project course. The videos 
were an addition to textual descriptions of scenarios in order to ease the 
communication between developers and customers. The scenario-based videos 
helped to resolve misunderstandings and ambiguities.
Brill et al. \cite{Brill.2010} analyzed the use of low-effort ad-hoc videos 
that show scenarios of a future system compared to textual use cases. Their 
results yielded that such videos helped to avoid misunderstandings and 
clarified requirements better than use cases in the early phases of a project.
Xu et al. \cite{Xu.2012} developed an approach of evolutionary scenario-based 
design which advocates the use of videos for scenarios to represent 
unimplemented parts of a system. These videos were used for requirements 
elicitation and system demonstrations to support effective communication. The 
authors presented an approach to simplify the video production and modification 
using virtual world technology.
The VisionCatcher of Pham et al. \cite{Pham.2012} supported the creation of a 
multimedia representation of visionary scenarios for a system under 
development. The representations could be easily created, modified and replayed 
in meetings with stakeholders to achieve a common ground.
Stangl et al. \cite{Stangl.2011, Stangl.2012} presented SCRIPT which is a 
framework to combine scenarios and prototyping to provide interactivity as well 
as traceability of requirements. Even though the major contribution of the 
framework is consistency between scenarios and prototypes, SCRIPT also supports 
the generation of videos. These videos were based on the predefined transitions 
between different mockups of a specified scenario.

The previously mentioned approaches already show that the combination of 
videos and scenarios support communication to achieve a \textit{shared 
understanding}. However, the production and modification of video is a major 
problem due to its high effort. Inspired by the approach of Stangl et al.
\cite{Stangl.2011,Stangl.2012}, we decided to focus on generating videos to 
document scenarios with respect to the dynamic aspect of interaction. Our idea 
is to integrate the video as a by-product of digital prototyping. In contrast 
to Stangl et al. \cite{Stangl.2011,Stangl.2012}, our approach does not require 
any predefined transitions between mockups and we can capture mouse, keyboard 
and touch events. Thus, we can easily define scenarios in meetings with 
stakeholders, similar to Pham et al. \cite{Pham.2012}. Our selected 
representation of interaction enables to replay the captured process at any 
time without previously generating videos. A generated video is just an 
exportable documentation option. Such a video is more suitable for 
communication and can be used by anyone independently from a software 
application that is implementing our approach.

\section{Video as a By-Product of Digital Prototyping}
\label{ch:approach}
Our approach for video as a by-product of digital prototyping is based on 
Schneider's \textit{by-product approach} \cite{Schneider.2006}. This concept is 
defined by two goals that should be achieved by seven principles. We had to 
adjust the \textit{by-product approach} slightly for our purpose of documenting 
interaction with video since Schneider \cite{Schneider.2006} focused on 
the rationale of requirements. The adjusted two goals and seven principles are 
as followed:\\
\textbf{Goals}
\begin{enumerate}
	\item Capture information to be documented by video during 
	specific tasks within software projects
	\item Be as little intrusive as possible to the bearer of the information 
	to be documented by video
\end{enumerate}
\textbf{Principles}
\begin{enumerate}
	\item Focus on a project task in which information to be documented by 
	video is surfacing
	\item Capture information to be documented by video during that task (not 
	as a separate activity)
	\item Put as little extra burden as possible on the bearer of the 
	information to be documented by video (but maybe on other people)
	\item Focus on recording during the original activity, defer indexing, 
	structuring etc. to a follow-up activity carried out by others
	\item Use a computer for recording and for capturing additional 
	task-specific information for structuring
	\item Analyze recordings, search for patterns
	\item Encourage, but do not insist on further management of information to 
	be documented by video
\end{enumerate}

These principles help to shift effort away from the respective project tasks 
(\textit{goal $1$}) and from the bearers of the information (\textit{goal $2$}) 
\cite{Schneider.2006}. In the following, we explain our concepts that consider 
the principles to achieve the goals.

\subsection{Support of Arbitrarily Created Mockups}
We focused on the dynamic aspect of interaction as the information to be 
documented by video. Prototyping is one specific task within a project 
(\textit{principle 1}) in which we can capture this information 
(\textit{principle 2}). However, mockups can be hand- or digitally drawn with 
different levels of visual refinement. We can support arbitrarily created 
mockups by digitalizing them if necessary and adding responsive controls with a 
user interface builder, e.g. the Gluon Scene Builder \cite{Gluon}. Thus, we 
apply digital prototyping, which requires the use of a computer 
(\textit{principle 5}). The creation of mockups and their overlay with 
responsive controls needs to be done before a prototyping session by the 
requirements engineer. Thus, there is no extra burden for the stakeholders, who 
are the bearers of the information (\textit{principle 3}). A stakeholder only 
needs to describe the desired interaction or even interact with the digitalized 
mockups.

\subsection{Evolutionary Scenario Specification}
A prototyping session can be used for analyzing how a user wants to 
interact with a system to elicit scenarios or following through predefined 
scenarios to obtain stakeholder feedback for modifications according to users' 
needs \cite{Stangl.2012}.

Both purposes require that scenarios can be created and modified fast and easy 
to evolve during their specification. Digital prototyping allows us to record 
the dynamic aspect of interaction for video documentation during the original 
activity itself (\textit{principle 4}). We use a computer for capturing 
specific information (\textit{principle 5}) that can be used to generate videos 
which demonstrate the interaction sequence of scenarios. This specific 
information for one scenario includes the order of the mockups and the 
interaction event sequence for each mockup. Any modification of a scenario 
sequence can be achieved by adding, rearranging and deleting mockups. Whereas 
added mockups require specifying the necessary interaction events, rearranged 
ones can maintain the specified interaction sequences. The interaction 
sequences can be edited for each mockup separately. There are two options: 
(1) deleting and recording a whole new interaction sequence or (2) 
capturing new single events and arranging them in the order of the existing 
ones.

\subsection{Video-Independence}
The recording of the selected information allows focusing on the original 
activity (\textit{principle 4}). From the collected data of the mockups' order 
and their interaction event sequences, we can simulate the specified scenarios 
at any time. Thus, we are independent of video which reduces effort and saves 
time since no video needs to be generated. A repeated analysis of the 
defined scenarios is necessary for reconsideration during a prototyping 
session and afterward (\textit{principle 6}). Since a generated video is only 
an export medium, we encourage the further management of the collected data 
but do not insist on it (\textit{principle 7}). The export of videos is 
an important factor to be independent of an application that implements our 
approach.\\

We considered all seven principles in the development of our three concepts. 
Thus, we could achieve the two defined goals for video as a by-product of 
digital prototyping.

\subsection{Software Tool: Mockup Recorder}
\begin{figure*}[htbp]
	\centering
	\includegraphics[width=1.0\linewidth]{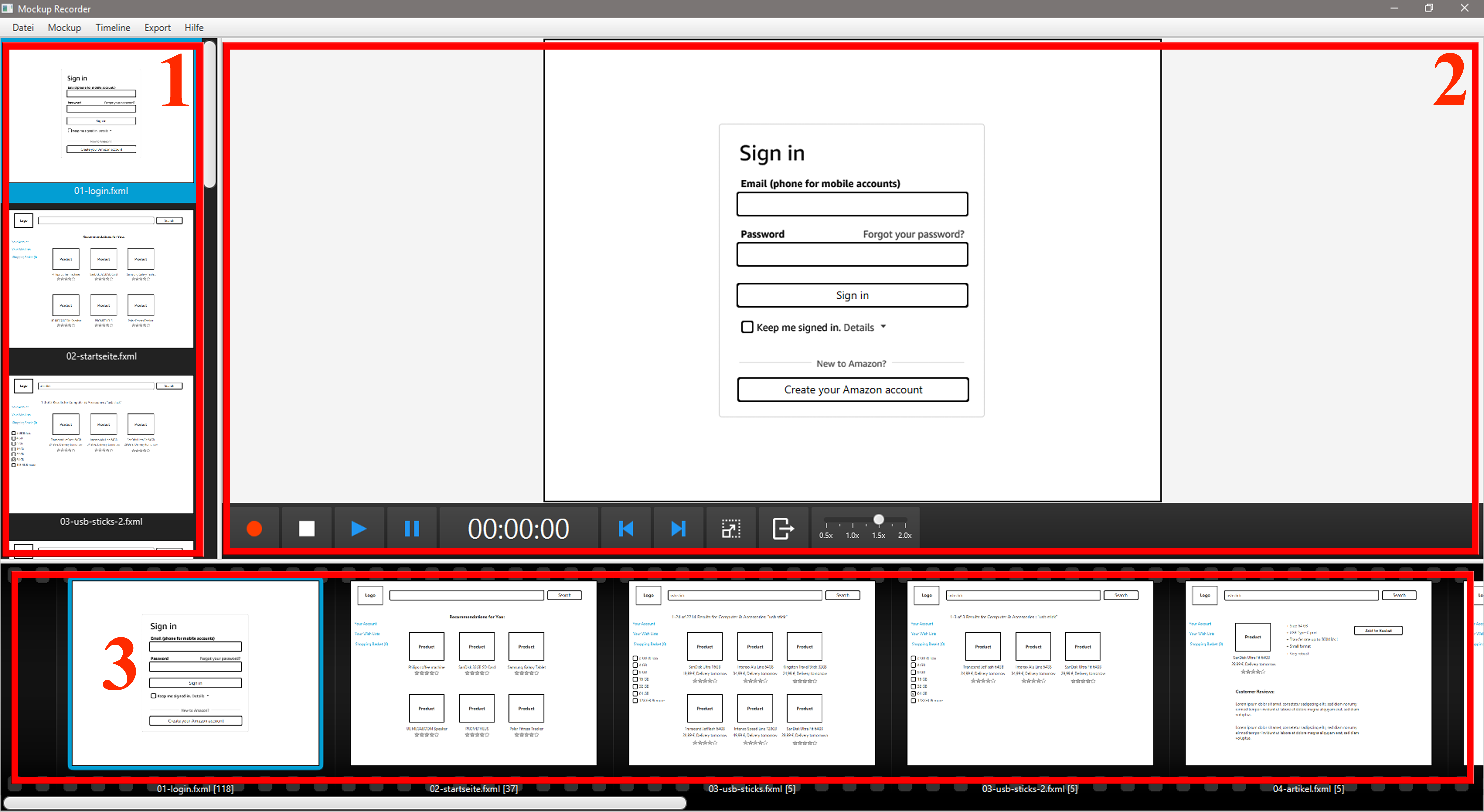}\hfill
	\caption{Mockup Recorder: 'Selection of existing mockups' (1), 
		'Mockup preview to capture and replay interaction' (2) and 'Timeline of 
		a scenario' (3)}
	\label{fig:tool}
\end{figure*}

We implemented all three concepts in a prototypical software tool called 
\textit{Mockup Recorder} (see \figurename{ \ref{fig:tool}}). This software tool 
allows to create and import mockups (see \figurename{ \ref{fig:tool}}, (1)). We 
support the use of hand-drawn and digitally created mockups by adding 
responsive controls. The mockups can be arranged on a timeline to define the 
sequence of a scenario (see \figurename{ \ref{fig:tool}, (3)}). We can modify 
such a sequence by adding, rearranging or deleting mockups within the timeline. 
We capture and replay interaction events of responsive controls by using the 
mockup preview (see \figurename{ \ref{fig:tool}}, (2)). The interaction event 
sequences are stored separately for each mockup in the timeline. When one 
sequence ends, we navigate to next mockup. Currently, any editing of the 
interaction events of a single mockup 
requires deleting all events of this mockup and recording them again. Defined 
scenarios can be exported as separate videos. We can also export the 
created and imported mockups as images.

\section{Evaluation}
\label{ch:evaluation}
Although our approach simplifies the production and modification of videos in 
comparison with screencasts, the benefit of our generated videos is so far 
unknown. Since the major documentation option for specifications is still the 
written natural language, we perceive both mockups and videos as additional 
support to understand textual descriptions. Therefore, the aim of our 
evaluation was to investigate whether a textual scenario can be faster 
understood with the support of a dynamic video or static mockups. We proceeded 
to achieve this objective by comparing the two different media as additional 
support for a textual scenario. This investigation allows us to judge whether 
our generated videos provide a benefit against static mockups. We asked the 
following research question:

\begin{itemize}[\settowidth{\labelwidth}{RQ:}]
\item[RQ:] Can a textual scenario be faster and better understood with the 
support of a dynamic video or static mockups?
\end{itemize}
We tested the following three null hypotheses:

\begin{itemize}[\settowidth{\labelwidth}{$H3_{0}$:}]
\item[$H1_{0}$:] There is no speed difference in familiarizing oneself with a 
textual scenario supported by either a dynamic video or respectively static 
mockups.

\item[$H2_{0}$:] There is no speed difference in extracting information from a 
textual scenario and its additional support in terms of either a dynamic video 
or respectively static mockups to answer questions.

\item[$H3_{0}$:] There is no difference in the number of correct answers based 
on the extracted information from a textual scenario and its additional support 
in terms of either a dynamic video or respectively static mockups.
\end{itemize}

Each corresponding alternative hypothesis $Hi_{1}, i \in \{1,2,3\}$ considers 
that the respective difference exists.

\begin{table*}[!b]
	\centering
	\caption{Experiment results -- Training time [s], process time [s], number 
		of correct answers}
	\label{experiment_results}
	\begin{tabular}{|c|c|c|c||c|c|c|c|}
		\hline
		\multicolumn{4}{|c||}{\textbf{Group: Video}} & 
		\multicolumn{4}{c|}{\textbf{Group: Mockups}} \\ \hline
		\textbf{Subject} & \textbf{Training Time} & \textbf{Process Time} & 
		\textbf{Correct Answers} & \textbf{Subject} & \textbf{Training Time} & 
		\textbf{Process Time} & \textbf{Correct Answers} \\ \hline \hline
		P1 & 124 & 160 & 6 & P9 & 225 & 245 & 9 \\ \hline
		P2 & 154 & 333 & 9 & P10 & 160 & 175 & 10 \\ \hline
		P3 & 140 & 190 & 8 & P11 & 171 & 160 & 9 \\ \hline
		P4 & 133 & 165 & 9 & P12 & 150 & 188 & 10 \\ \hline
		P5 & 85 & 180 & 9 & P13 & 96 & 232 & 8 \\ \hline
		P6 & 144 & 251 & 9 & P14 & 128 & 228 & 10 \\ \hline
		P7 & 90 & 204 & 8 & P15 & 273 & 250 & 9 \\ \hline
		P8 & 90 & 175 & 10 & P16 & 237 & 173 & 10 \\ \hline \hline
		Mdn & 128.5 & 185.0 & 9.0 & Mdn & 165.5 & 208.0 & 9.5 \\ \hline
		SD & 27.63 & 58.35 & 1.20 & SD & 59.84 & 36.07 & 0.74 \\ \hline
	\end{tabular}
\end{table*}

\subsection{Evaluation Design}
In this evaluation, we performed a between-subjects experiment with two groups. 
Whereas the group \textit{video} got a video as additional support, the 
group \textit{mockups} got the corresponding $11$ static mockups. We measured 
three dependent variables: The \textit{training time} to familiarize oneself 
with the given material, the \textit{process time} to answer questions by 
extracting information from the given material and the number of 
\textit{correct answers}. The independent variable was the additional material 
for a textual scenario with two levels: a video and the static mockups. The 
textual scenario consisted of $19$ steps. If a step initiated an interaction 
event sequence, we added a reference from the step to the respective mockup. We 
measured the \textit{training time} and \textit{process time} with a 
stopwatch and the number of correct answers by using a questionnaire. The 
experiment represents a scenario in which the subject is a developer who has to 
understand a scenario of how a customer wants to buy a product in a web store 
in order to implement the corresponding software. We focused on the perspective 
of a developer since this role mainly works with the artifacts of the 
requirements analysis.

\subsection{Evaluation Procedure}
The experiment was carried out within one week with $16$ subjects consisting of 
$14$ undergraduate and $2$ graduate students of computer science. 
All subjects had a similar level of knowledge with respect to scenarios and 
mockups as well as at least one year experience as a developer. We randomly 
assigned the subjects to one of the two groups. Regarding 
the random assignment, we only ensured that the undergraduate and graduate 
students were equally distributed to both groups. A session with one subject 
included an introduction to the experiment with its two tasks of familiarizing 
oneself with the given material and subsequently extracting information from it 
to answer questions. We performed the two tasks one after the other to measure 
the \textit{training time} and \textit{process time} separately. For the first 
task, we measured the \textit{training time} from the beginning of the task 
until the subject explicitly stated to be familiar with the material. For the 
second task, we measured the time from the beginning of the task until the 
subject answered all $10$ questions of the questionnaire. These questions 
focused on detailed aspects of the given scenario. For example, we asked for 
presented information such as delivery options or specific steps of the 
interaction process itself like the order of how a customer enters the data for 
a delivery. We permitted the subjects to use the given material for answering 
the questions since we wanted to know how the subjects work with the artifacts 
and not how much they can memorize.

\subsection{Analysis and Results}
\tablename{ \ref{experiment_results}} shows the results of each subject for the 
respective group of our experiment. For each of the three dependent variables, 
we performed an independent 2-group Mann-Whitney U test at a significance level 
of $p = 0.05$. Thus, we can determine whether an observed difference between 
the two groups exists due to the test conditions or by chance. In case of an 
observed difference, we additionally calculated Cohen's $d$ and the statistical 
power $1 - \beta$. Cohen's $d$ is the most common type of effect size to judge 
whether or not the difference between two groups' mean is large enough to have 
practical relevance. The statistical power $1 - \beta$ is the probability of 
the correct decision to reject the null hypothesis $Hi_{0}$ if the alternative 
hypothesis $Hi_{1}, i \in \{1,2,3\}$ is true.

\tablename{ \ref{analysis_results}} presents the results of the conducted 
independent 2-group Mann-Whitney U tests.
The first Mann-Whitney U test indicated that the \textit{training time} to 
familiarize oneself with the given material was significantly shorter for video 
($Mdn = 128.5 s$) than for static mockups ($Mdn = 165.5 s$) as additional 
support, $U = 10, p = 0.024$. Hence, $H1_{0}$ can be rejected. Video as 
additional support shortens time to familiarize oneself with a textual scenario 
compared to mockups. The value of Cohen's $d$ is $1.287$ and thus greater than 
the threshold of $0.8$ for a large effect \cite{Cohen.1992}. The identified 
difference between video and mockups as additional support for a textual 
scenario has practical relevance. The statistical power $1 - \beta$ is as 
much as $0.739$, which is close to the required threshold of $0.8$ proposed by 
Cohen \cite{Cohen.2009}. Thus, we are optimistic of rejecting $H1_{0}$ and 
accepting $H1_{1}$. The validity of this result, however, is restricted.
The second Mann-Whitney U test showed no difference in the \textit{process 
time} to extract information from the given material to answer questions 
between video ($Mdn = 185.0 s$) and static mockups ($Mdn = 208.0 s$), 
$U = 31, p = 0.958$. The null hypothesis $H2_{0}$ cannot be rejected. Cohen's 
$d$ and the statistical power $1 - \beta$ cannot be calculated due to 
a missing difference.
The third Mann-Whitney U test yielded no difference in the number of 
\textit{correct answers} based on the extract information of a textual scenario 
supported by video ($Mdn = 9.0$) or by static mockups ($Mdn = 9.5$), $U = 17, p 
= 0.105$. Consequently, we cannot reject $H3_{0}$. Hence, Cohen's $d$ and the 
statistical power $1 - \beta$ cannot be determined.

\begin{table}[htbp]
	\centering
	\caption{Independent 2-group Mann-Whitney U Test}
	\label{analysis_results}
	\begin{tabular}{|c||c||c||c|}
		\hline
		\textbf{DV} & \textbf{Training time} & \textbf{Process time} & 
		\textbf{Correct answers} \\ \hline 
		\hline
		$U\textnormal{-}value$ & 10 & 31 & 17 \\ \hline
		$p\textnormal{-}value$ & 0.024 & 0.958 & 0.105 \\ \hline
		Cohen's $d$ & 1.287 & -- & -- \\ \hline
		Stat. Power & \multirow{2}{*}{0.739} & \multirow{2}{*}{--} & 
		\multirow{2}{*}{--} \\
		$1 - \beta$ &  &  &  \\ \hline
	\end{tabular}
\end{table}

\subsection{Interpretation}
Our findings provide insights with respect to the benefit of a dynamic video as 
additional support of a textual scenario compared to static mockups.

Whereas video statistically significantly shortens time to familiarize oneself 
with a textual scenario, we could not find any speed difference in extracting 
information from the given material to answer questions. There was also no 
difference in the number of correct answers between both supporting media.

The use of our generated video ($Mdn = 128.5 s$) leads to $22.36 \%$ less 
\textit{training time} than the corresponding mockups ($Mdn = 165.5 s$) before 
being familiar with a textual scenario. Thus, video enables developers to 
get a faster understanding of a scenario than mockups. This finding has 
practical relevance which the corresponding effect size emphasizes. 
Additionally, a video is as suitable as mockups to extract information and 
answer questions with respect to the content and interaction process of a 
scenario. Hence, the understanding of a textual scenario with the support of a 
video is as good as with mockups. As an answer to our research question, we can 
summarize:

\begin{itemize}[\settowidth{\labelwidth}{A:}]
	\item[A:] A textual scenario can be faster understood with the support of a 
	dynamic video, generated by our approach, than with the support of static 
	mockups. Both additional media lead to an equally good understanding. Video 
	allows capturing the dynamic aspect of interaction and provides developers 
	the benefit of 	familiarizing themselves faster with a scenario. Thereby, 
	the extraction of information is as good as using static mockups.
\end{itemize}

\subsection{Threats to Validity}
In the presented evaluation, we considered threats to validity corresponding to 
the classification of Wohlin et al. \cite{Wohlin.2012}.

\subsubsection{Construct Validity}
We have a mono-operation bias since we only selected one exemplary scenario of 
how a customer wants to buy a product in a web store. As a consequence, our 
evaluation does not convey a comprehensive representation of the real world 
complexity. Nevertheless, our selected scenario represents one challenging 
situation of the real world. In the evaluation, we only used objective 
measures, which is a mono-method bias. This threat to validity only allows a 
restrict explanation of our findings. However, we decided to focus on objective 
measures since they can be reproduced more easily and are thus more reliable 
than subjective ones. The second task of extracting information to answer 
questions caused an interaction of testing and treatment. The answering of 
questions implies to measure the number of correct answers. Therefore, our 
subjects could be more aware of their errors as a factor. Maybe this influenced 
the process time of the respective groups since the subjects could have taken 
more time to answer the questions than necessary.

\subsubsection{Internal Validity}
We had two different groups due to the selected between-subjects design for the 
evaluation. These groups caused interactions with selection since different 
groups have a different behavior. However, we consciously decided to use this 
evaluation design in order to use only one textual scenario in both groups. 
Thus, we counteracted learning effects and achieved a better comparability. 
The distribution of the participants over one week is a further threat to 
internal validity. The respective daytime could have had an influence on the 
subjects and their motivation to contribute to our evaluation.

\subsubsection{Conclusion Validity}
We decided to use objective measures to increase the reliability of our 
measures. Objective measures are easier to reproduce and more reliable than 
subjective ones. However, we determine the number of correct answers by using a 
questionnaire. A poor question wording could have an influence on subjects' 
understanding. Therefore, we allowed the subjects to ask questions in case of 
ambiguity. The calculated statistical power of the identified difference in the 
training time is below the required threshold of $0.8$ according to Cohen 
\cite{Cohen.2009}. Even though the statistical power is $0.739$, our results 
are currently not sufficient to ensure that we did not draw erroneous 
conclusions. All subjects had at least one year experience as a developer which 
makes them a homogeneous group to counteract the threat of erroneous 
conclusions. Thus, we mitigated the risk that the variation due to the 
subjects' random heterogeneity is larger than due to the investigated 
supporting media for the textual scenario.

\subsubsection{External Validity}
All subjects are representative of our target group of developers since all of 
them had at least one year experience as a developer. However, the experimental 
setting endangered the external validity since the environment was different 
from the real world. The selected scenario of how a customer wants to buy a 
product in a web store had no pragmatic value for the subjects. None of them 
had a genuine working task with given material of the selected scenario. This 
scenario is also a general one that probably all subjects have experienced. 
Their prior knowledge could have had an influence on their answers. We 
counteracted this threat to validity by changing steps in the interaction 
process and the presented data of the web store scenario.

\section{Discussion}
\label{ch:discussion}
This paper proposes an approach to capture the dynamic aspect of interaction by 
video.

Although there are practices to analyze at an early stage how a user wants to 
interact with a future software, users are often dissatisfied with the 
resulting usability. Scenarios are one effective requirements engineering 
practice to reveal important aspects of interaction. Natural language is the 
most common notation to document scenarios. This textual documentation option 
is static and thus often improper to capture and communicate knowledge about 
interactions. For several years, video is known as the best documentation 
option for communication. However, videos are currently not an established part 
of requirements engineering. There is a need for tools and methods to integrate 
videos in the existing activities.

Our approach consists of three concepts to achieve video as a by-product of 
digital prototyping. Videos as a by-product can help to reduce the effort of 
video production and thus lower the threshold for using videos to make them 
more attractive for practitioners and researchers. 

Developers can familiarize themselves faster with a textual scenario supported 
by a generated video from our approach, than by the corresponding mockups. We 
assume that watching a video and reading a text in arbitrary order can be done 
faster than switching between mockups and a text to match these two artifacts. 
Furthermore, we identified no difference in the extraction of information from 
a textual scenario and the respective supporting media regarding speed and 
number of correct answers. Thus, a video is as good as mockups to enrich a 
textual scenario for identifying specific information. Both supporting media 
lead to a similar extraction effort and equally good results.

All in all, our generated video leads to an equally good but statistically 
significant faster understanding of a textual scenario than mockups. The 
practical relevance of our finding is substantiated by the determined effect 
size. Our approach does not require any additional effort to produce videos 
since we integrated video generation in digital prototyping. Thus, we achieved 
the rarely used but best documentation option ``video'' as a by-product. 

A generated video of our approach is a stable reference for post-processing 
work. It represents a strictly defined sequence of a scenario. This fixed 
documentation option is a benefit in comparison with the bare interactive 
prototype since no interaction by a user is necessary. Thus, a scenario can be 
played repeatedly without any deviation from its original sequence. 
Furthermore, a video can be easily shared and used by anyone since no 
additional knowledge or software tool is necessary. These advantages of the 
documentation option ``video'' contribute to a \textit{shared understanding} 
between a project team and its stakeholders and help to improve the 
\textit{specification quality} of user requirements with respect to interaction.

Currently, we are experimenting with the generated videos. We want to achieve a 
more comprehensive insight into the videos' potential for requirements 
communication to bridge the communication gap between a project team and its 
stakeholders.

\section{Conclusion}
\label{ch:conclusion}
This work contributes an approach to specify and capture the dynamic aspect of 
interaction by video. Requirements engineering provides several practices, such 
as scenarios, to reveal important aspects of interaction. However, their static 
description with natural language is often improper to document and communicate 
interaction knowledge.

We propose an approach consisting of the three concepts \textit{support of 
arbitrarily created mockups}, \textit{evolutionary scenario specification}, and 
\textit{video-independence} to integrate the best known but rarely used 
documentation option ``video'' as a by-product of digital prototyping. Thus, we 
achieve an easy-to-modify and always repeatedly playable representation of 
interaction. This data can be used to generate videos that document the 
interaction sequence of scenarios. We implement these concepts in the 
prototypical software tool \textit{Mockup Recorder}. A first evaluation showed 
that a textual scenario supported by our generated video can be faster and 
equally good understood compared to the static mockups.

Our work points to the conclusion that videos are a suitable documentation 
option as additional support for a textual scenario. Videos as a 
by-product of existing requirements engineering activities can support to 
achieve the two major goals of requirements engineering: \textit{shared 
understanding} and \textit{specification quality}. Thus, our approach 
contributes to requirements engineering success which is important to 
achieve a software product with satisfying \textit{usability}.

\section*{Acknowledgment}
This work was supported by the German Research Foundation (DFG) under ViViReq 
($2017$ -- $2019$).

\bibliographystyle{IEEEtran}
\bibliography{IEEEabrv,ref}

\end{document}